\documentclass[final,3p,times,sort&compress]{elsarticle}
\usepackage{color}
\usepackage{url}
\usepackage{multirow,setspace,times,amssymb,amsmath,graphicx,color,subfigure,url}
\usepackage{dsfont}
\usepackage{threeparttable}
\usepackage{rotating}

\usepackage{dcolumn}
\usepackage{listings}

\journal{Elsevier}

\begin{document}

\begin{frontmatter}

\title{The short-term effect of COVID-19 pandemic on China's crude oil futures market: A study based on multifractal analysis}

\author[SUIBE]{Ying-Hui Shao}

\author[SLUAF]{Ying-Lin Liu }

\author[ECNU]{Yan-Hong Yang \corref{cor1}}
\ead{yhyang@dedu.ecnu.edu.cn}

\cortext[cor1]{Corresponding author.}

\address[SUIBE]{School of Statistics and Information, Shanghai University of International Business and Economics, Shanghai 201620, China}
\address[SLUAF]{School of Financial Technology, Shanghai Lixin University of Accounting and Finance, Shanghai 201209, China}
\address[ECNU]{Faculty of Education, East China Normal University, Shanghai 200062, China}

\begin{abstract}
 The ongoing COVID-19 shocked financial markets globally, including China's crude oil future market, which is the third most traded crude oil futures after WTI and Brent. As China's first crude oil futures accessible to foreign investors, the Shanghai crude oil futures (SC) have attracted significant interest since launch at the Shanghai International Energy Exchange. The impact of COVID-19 on the new crude oil futures is an important issue for investors and policy makers. Therefore this paper studies the short-term influence of COVID-19 pandemic on SC via multifractal analysis. We compare market efficiency of SC before and during the pandemic with the multifractal detrended fluctuation analysis and other commonly-used random walk tests. Then we generate shuffled and surrogate data to investigate the components of multifractal nature in SC. And we examine cross-correlations between SC returns and other financial assets returns as well as SC trading volume changes by the multifractal detrended cross-correlation analysis. The results show that market efficiency of SC and its cross-correlations with other assets increase significantly after the outbreak of COVID-19. Besides that, the sources of its multifractal nature have changed since the pandemic. The findings provide evidence for the short-term impacts of COVID-19 on SC. The results may have important implications for assets allocation, investment strategies and risk monitoring.
\end{abstract}
\end{frontmatter}

\section{Introduction}
\label{s:introduction}
The Shanghai crude oil futures (SC) have received extensive attention among practitioners and academics recently. The Shanghai International Energy Exchange (INE) launched SC on March 26, 2018 to strengthen the crude oil pricing power, which is China's first crude oil futures contract open to foreign investors. Almost half of China's oil imports are originated in the Middle East, which produces sour crude containing between 0.5\% and 2.0\% of sulfur by weight. Accordingly the SC futures contract is based on medium sour crude oil with 1.5\% of sulfur. Besides that, SC has a different quotation currency from other international crude oil futures such as West Texas Intermediate (WTI), which is dollars-measured. To support the internationalisation of the Renminbi (RMB), daily settlement variation and physical delivery settlement of SC are denominated in RMB.

With rapid development, SC is conducive to reflecting supply and demand conditions in the Asia-Pacific region. SC provides investors another instrument to trade the market and has been the third most traded crude oil futures contract after WTI and Brent, which are the two major global crude oil benchmarks. The daily trading volume of SC exceeded 200,000 which is equivalent to that of WTI and Brent in Asian trading time. And SC also has been attracting a lot of interest from academics. There are a growing body of literature on SC, including market efficiency \citep{yang2020pricing}, volatility forecasting \citep{lu2020examining}, return and volatility spillover \citep{zhang2021exploring}, and price-volume relationship \citep{zhang2021cross}. 

In addition, crude oil futures market is highly fragile to extreme events \citep{yang2020time,shao2020does} and SC is no exception. Since the Wuhan Municipal Health Commission reported a cluster of cases of pneumonia in Wuhan at the end of 2019, the ongoing COVID-19 pandemic has not only caused a worldwide devastating health crisis, but also massive disruptions to global economy. The overwhelming stresses created by COVID-19 pandemic have spread to almost every sector of the world, resulting in great losses especially in banks, consumer discretionary, energy, and industrials. The rapid spread of COVID-19 shocked financial markets all over the world, including the China's crude oil futures market. China adopted a number of measures to prevent the pandemic spread, including lockdown, travel restriction, wearing mask and social distancing. INE also adjusted price limit and trading schedule  for risk management. These measures had a profound and systematic impact on the commodity markets, economy activity and foreign trade. Even though there is a rapid growth in trading volume and research interest of SC, it is still under development and not fully investigated. There is an urgent need to uncover how the COVID-19 affects SC. 

In this paper, we utilize the multifractal detrended fluctuation analysis (MF-DFA) \citep{kantelhardt2002multifractal} and the multifractal detrended cross-correlation analysis (MF-DCCA) \citep{zhou2008multifractal} to investigate the short-term impact of COVID-19 on SC, which are popular multifractal analysis techniques. Multifractality is commonly reported in most financial assets and significant for predictability, market efficiency test and risk management \citep{2019multifractal}. The multifractal origins are fundamental to the behavior and mechanisms of financial assets. Thus we comprehensively compare the market efficiency and multifractal sources of SC before and during the new coronavirus. Then we investigate the cross-correlations between SC and other financial assets returns as well as SC's trading volume changes via the MF-DCCA. 

Our study contributes to the current literature on several fronts. First, the  comparative studies of SC before and during the COVID-19 pandemic have been scarcely reported. The behaviors of this new comer, including market efficiency and price-volume relationship are still open to question. This paper provides fresh insights of the COVID-19 influence on this market based on multifractal analysis. Second, the sample period is two months before to two months after the outbreak of COVID-19. Measures to boost the economy and slowdown the spread of the pandemic have not yet completely been realised. This allows us to understand how the COVID-19 affects SC during a very stressful time. Third, this paper investigates nonlinear linkage between this newly-shipped crude oil futures and other financial assets. Efficient portfolio management requires a deep understanding of the linkages among different assets. Our findings have potential applications in assets allocation, portfolio construction and risk management.

Our analysis provides evidence of the impact introduced by the COVID-19 on SC. Interestingly, market efficiency of SC increases after the outbreak of COVID-19. Cross-correlations between SC and other assets also improve during this period, which indicates the presence of systemic risk.  

The rest of this study is organized as follows. Section \ref{s:literature} presents a literature review of relevant papers. Section \ref{s:data} depicts the data. Section \ref{s:method} introduces the methodologies. Section \ref{s:results} presents emprical results. Section \ref{s:conclusion} concludes the paper.

\section{Literature review}
\label{s:literature} 
A number of studies examine the information efficiency of crude oil market. Empirical results reveal that WTI and Brent are weak-form efficient in the long run, though exhibit inefficiency over a short term \citep{kristoufek2014commodity,ghazani2019testing,shao2020does}, which implies the behavior of crude oil market might be predictable at a short time scale. Similarly, SC is close to weak-form efficient from March 2018 to February 2019 \citep{yang2020pricing}, which is a relatively long period. There is little study on SC market efficiency over a shorter time period.

Much of the existing literature explores the relations and differences between the nascent SC and other crude oil markets, especially WTI and Brent (the two international crude oil benchmarks). There exist cointegration relationships and significant dynamic interactions among SC and other major crude oil futures markets \citep{yang2020return}. The equilibrium relationship between SC and other spot markets is also documented \citep{yang2020pricing}. Multifractal characteristic of SC is stronger than WTI and weaker than Brent \citep{wang2019multifractal}. Compared with the interaction between WTI and Brent, the comovement strength between SC and the two international crude oil futures is weaker and more unstable \citep{huang2020identifying}. Moreover, SC displays different interactions with international benchmarks during the day and night trading sessions \citep{ji2021intra}.

The relationship between crude oil futures and stock markets also has received considerable critical attention. Several current literature investigates the nexus and portfolios between SC and stock markets. It is verified that there are notable industry characteristics in this relationship \citep{zhu2021relationships}. Most of the Chinese stock sectors exhibit higher upper tail dependence with SC than WTI \citep{zhu2021relationships}. Other study provides contrary evidence that SC is a better choice for portfolio diversification. Compared with WTI, SC results in a better diversification effect and volatility reduction in multi-asset allocation with Chinese petrochemical-related stocks \citep{lv2020crude}. 

Moreover, co-movement between crude oil prices and exchange rates is of great significance to monetary policy-making and portfolio management strategy. Researchers find there is close interaction between the oil prices in domestic currencies and the dollar exchange rates of those currencies \citep{mensi2017oil}. And there is different dependent structure between foreign exchanges and crude oil futures for various countries and markets \citep{mensi2017oil,ahmad2020intraday}. Exchange rate has a weak driving effects on SC during the COVID-19 pandemic \citep{lin2021china}. Since RMB exchange rates in onshore and offshore markets plays an important role in futures trading, we also explore the cross-correlations between SC and the two RMB exchange rates. 

Recently, the ongoing COVID-19 has triggered uncertainty globally. During this pandemic, there are stronger spillovers from stock markets to crude oil markets and significant stock-SC risk spillovers \citep{zhu2021multidimensional}. And the importance of Chinese financial factors as SC price drivers increased considerably than before \citep{lin2021china}. Global coronavirus news is a key factor for the China's crude oil volatility prediction \citep{niu2021role}. The return connectedness among commodity assets including SC and financial assets increase rapidly after the outbreak of COVID-19 \citep{li2021return}. 

Up to now, limited attention has been paid to the nonlinear structure of SC during this black-swan like event. Our study based on multifractal analysis complement the existing literature on SC. 

\section{Data}
\label{s:data}
We utilize 5-minute closing price from the Wind database of the following categories: (i) continuous futures contracts of SC, WTI and Brent, (ii) the Chinese stock index (CSI300), (iii) onshore and offshore RMB exchange rates. Our dataset also includes the trading volume of SC. We compute the first difference of the natural logarithm of price as returns and calculate volume changes analogously. Given the time zone difference between China and the US, we synchronize the time of each US financial instrument to Beijing Time (UTC+8) with daylight savings adjustments.  

On December 31, 2019 Wuhan Municipal Health Commission reported a pneumonia outbreak in the City, which is the first COVID-19 official report globally. In this study, we mainly focus on the short-term effects of COVID-19 on SC. Thus pre-pandemic spans from November 2019 to December 2019 and period between January 2020 and February 2020 corresponds to the COVID-19 pandemic.

\section{Methodology}
\label{s:method}
\subsection{Random walk tests} 
In a weak-form efficient market, prices are not predictable due to the random nature of future events. Correspondingly, one can employ statistical tests for random walk hypothesis to examine market efficiency. If the tests reject the random walk hypothesis, the market is not weak-form efficient. We conduct (1) Runs test with null hypothesis that data comes in random order \citep{wald1940test}, (2) Ljung-Box test \citep{ljung1978measure} with the null hypothesis of no autocorrelation, (3) Variance Ratio test \citep{lo1988stock}, (4) BDS test \citep{broock1996test} which detects serial dependence in time series, (5) Mann–Kendall test \citep{mann1945nonparametric,kendall1975} which is used to identify an increasing or decreasing trend in dataset, and (6) detrended fluctuation analysis (DFA) \citep{peng1994mosaic}. We select the above mentioned approaches for reliability and validity.

Apart from the random walk tests, we utilize multifractal measurements to compare market efficiency of SC in different periods. And we study cross-correlations between SC and other financial time series with MF-DCCA \citep{zhou2008multifractal}. We represent details of DFA, MF-DFA and MF-DCCA in the following sections. 

\subsection{Multifractal analysis method}
We describe steps of MF-DCCA \citep{zhou2008multifractal} as follows. For two time series $\left\lbrace X_i, i=1,2,...,N\right\rbrace$ and $\left\lbrace Y_i, i=1,2,...,N\right\rbrace$, we split each into $N_{s} =$ int$\left[N/s\right]$ non-overlapping segments of size $s$. Then we calculate the cross-correlation function of each box as
\begin{equation}
\left [F_v(s)\right ]^{2}= \frac{1}{s}\sum_{i=1}^{s}\left[X_{v}(k)-\widetilde{X}_v(k)\right]\left[Y_{v}(k)-\widetilde{Y}_v(k)\right],
\end{equation}
where $\widetilde{X}_v(k)$ and $\widetilde{Y}_v(k)$ are the local trends in segments $\{{X}_v(k)\}_{k=1}^s$ and $\{{Y}_v(k)\}_{k=1}^s$ respectively. $\left [F_v(s)\right ]^{2}$, also called fluctuation function, can be understood as the detrended covariance between each pair of segments. There are a variety of approaches to calculate the local trends. Polynomial fitting is a common choice for estimation of $\widetilde{X}_v(k)$ and $\widetilde{Y}_v(k)$. In this instance, MF-DCCA corresponds to MF-X-DFA (multifractal detrended fluctuation cross-correlation analysis) \citep{zhou2008multifractal}, a MF-DCCA method based on DFA. In this study, MF-X-DFA is chosen since MF-X-DFA has been widely used in financial time series analysis \citep{2019multifractal}. For brevity we refer to MF-X-DFA as MF-DCCA.

We compute the overall $q$th order cross-correlation function as
\begin{equation}
\left\{
\begin{array}{ll}
F_{xy}(q,s)= \left[
\frac{1}{N_s}\sum_{v=1}^{N_s} \left|F_v(s)\right|^q\right]^{1/q},  q\neq0,\\

F_{xy}(0,s)={\rm exp}\left[\frac{1}{N_s}\sum_{v=1}^{N_s}\ln\left|F_v(s)\right| \right ], q=0.
\end{array}
\right.
\end{equation}

Varying box size $s$ and repeating the abovementioned process, we can determine the following power-law relationship between $F_{xy}(q,s)$ and $s$
\begin{equation}
F_{xy}(q,s)\sim s^{h_{xy}(q)}
\label{Eq:MFDCCA:Fq_hq},
\end{equation}
where the scaling exponent $h_{xy}(q)$ is an estimation of generalized bivariate Hurst exponent. If $h_{xy}(2)\approx0.5$, the two time series are short-range cross-correlated. Otherwise the cross-correlations are long-range \citep{podobnik2011statistical}. We use the joint mass exponent function $\tau_{xy}(q)$ to characterize the joint multifractal nature via
\begin{equation}
\tau_{xy}(q)=qh_{xy}(q)-D_{f},
\label{Eq:MFDFA_tau1}
\end{equation}
where $D_{f}$ is the fractal dimension of the geometric support of the multifractal measure. We have $D_{f}$ = 1 for time series analysis. 
We calculate the joint singularity strength $\alpha_{xy}(q)$ and joint multifractal spectrum $f_{xy}(\alpha)$ through the Legendre transform \citep{halsey1986fractal}
\begin{equation}
\left\{
\begin{array}{lr}
\alpha_{xy}(q)=\rm{d}\tau_{xy}(q)/\rm{d}q,  \\
f_{xy}(\alpha)=q[\alpha_{xy}-h_{xy}(q)]+1.  
\end{array}
\right.
\label{Eq:MFDCCA_fa_a}
\end{equation}

We define $\Delta\alpha_{xy}$ as the difference between the maximum singularity $\alpha_{xy} (q_{\max})$ and the minimum singularity $\alpha_{xy} (q_{\min})$, which is the width of the joint singularity spectrum 
\begin{equation}
\Delta\alpha_{xy}=
\alpha_{xy} (q_{\max})-\alpha_{xy} (q_{\min}).
\label{Eq:MFDCCA_da}
\end{equation}
The joint singularity spectrum based measure $\Delta\alpha_{xy}$ is most widely adopted to quantify the joint multifractal strength. Usually one adopts a symmetric interval of $q$ with $q_{\max}+ q_{\min}=0$.

Similarly, we obtain another multifractal cross-correlation measure 
based on the scaling exponent $h_{xy}$, namely $\Delta h_{xy}$:
\begin{equation}
\Delta h_{xy}=
h_{xy} (q_{\max})-h_{xy} (q_{\min}).  
\label{Eq:MFDCCA_dh}
\end{equation}
If one observes bigger $\Delta\alpha_{xy}$ and $\Delta h_{xy}$ between two series than the other pairs, one can conclude that this pair of time series have higher joint multifractal strength. It also implies stronger multifractal cross-correlations. 

While $X = Y$, MF-DCCA reduces to MF-DFA, which is one of the most frequently used techniques to analyze multifractal nature in a single time series \citep{zhou2008multifractal,2019multifractal,shao2021Fractals}. In this case, $h(q)$ represents the generalized Hurst exponent. $\Delta\alpha$ and $\Delta h$ can be used to determine the strength of multifractality in a single time series. 

\subsection{Multifractal market efficiency measurements}
\label{s1}
With $q=2$ and $X=Y$, MF-DFA degrades to DFA, which is a widely used method to quantify long-range correlation in time series \citep{kantelhardt2002multifractal,Shao-Gu-Jiang-Zhou-Sornette-2012-SR, shao2015effects}. In this case, the scaling exponent $h(2)$ represents the Hurst exponent with range from 0 to 1, which is a popular market efficiency indicator \citep{lim2011evolution,2019multifractal,yang2019revisiting, shao2020does}. $h(2)\approx 0.5$ suggests a random-walk process. Otherwise, the time series exhibit long-range correlation. In this paper, we employ DFA to test weak-form efficient market  hypothesis. 

Multifractal strength can also serve to examine the degree of market inefficiency \citep{zunino2008multifractal}. Financial market with higher multifractality has lower market efficiency and higher risk \citep{zunino2008multifractal}. In this study, we start with random walk tests. If all the tests reject the null hypothesis, SC is not weak-form efficient. Then we use $\Delta \alpha$ and $\Delta h$ obtained by MF-DFA to compare market efficiency before and during the COVID-19. Additionally, we also adopt market deficiency measure ($MDM$) \citep{wang2009analysis,shao2021Fractals} to investigate information inefficiency of SC:
\begin{equation}
MDM=\frac{1}{2}\left( \left|h(q_{\min})-0.5\right|+\left|h(q_{\max})-0.5\right| \right) .
\label{Eq:MDM}
\end{equation}
Similar to $\Delta \alpha$ and $\Delta h$, higher $MDM$ implies lower market efficiency. 

\subsection{DCCA coefficient and corresponding statistical test}
In the case of $q=2$, MF-DCCA reduces to detrended cross-correlation analysis method (DCCA) \citep{podobnik2008detrended,zhou2008multifractal}. The multiscale detrended cross-correlation coefficient (DCCA coefficient) is defined as 
\citep{zebende2011dcca}: 

\begin{equation}
\rho_{DCCA}(s)=\frac{F^2_{xy}(s)}{F_x(s)F_y(s)},
\label{Eq:rho}
\end{equation}
where $F_x(s)$ and $F_y(s)$ are fluctuation functions of time series $X$ and $Y$ from DFA respectively. And $F^2_{xy}(s)$ is the detrended covariance function between $X$ and $Y$ from DCCA.  $\rho_{DCCA}(s)$, with range from -1 to 1, can be used to determine the cross-correlations between the nonstationary time series \citep{wang2013random}. $\rho_{DCCA}(s)=1$ suggests there is perfect positive  cross-correlation. Datasets with $\rho_{DCCA}(s)=-1$ exhibit perfect negative cross-correlation. When $\rho_{DCCA}(s)=0$, there are no cross-correlations. 

Following \cite{podobnik2011statistical}, we use $\rho_{DCCA}(s)$ to test whether the cross-correlations are statistically significant between two time series. We take a confidence level of 95\% and the null hypothesis of no cross-correlation, then generate i.i.d. normal distributed time series pairs without cross-correlations and determine the DCCA coefficient as critical values $\rho_{c}$. If $\rho_{DCCA}$ of real data is higher than $\rho_{c}$, the null hypothesis is rejected. Then we adopt the MF-DCCA to analyze the joint multifractal characteristics between SC and other financial time series. Otherwise, there is no significant cross-correlation between the two series.

\section{Results}
\label{s:results}
Extending work of \cite{yang2020pricing}, we run a battery of random walk tests to examine the weak-form market efficiency of SC pre-and during the COVID-19 pandemic. We report $p$-values of all the tests except DFA in Table~\ref{table:test}. And we present the Hurst exponent of DFA instead. Before the COVID-19, SC is not weak-form information efficient, since almost all the random walk tests apart from Mann–Kendall Rank test reject the null hypothesis  at 5\% significant level. Hurst index also indicates long-range dependence in SC. During the pandemic, most tests fail to reject the null hypothesis except Variance Ratio test and BDS test. And Hurst index is approximately equivalent to 0.5, which indicates a property of short-range dependence in SC. These results suggest that SC is close to weak form efficient during this period. 

\begin{table}[h]
	\renewcommand\arraystretch{1.4}
	\footnotesize
	\medskip
	\centering
	\caption{Results of weak form market efficiency tests.}
	\label{table:test}
	\begin{threeparttable}
		\begin{tabular}{cccccccc}   
			\hline
			&  &Runs & Ljung–Box &Variance Ratio& BDS& Mann–Kendall Rank &DFA 
			\\     
			\cline{1-8}
			& before &0.0001&0.0000&0.0000&0.0000&$0.7730^{*}$&0.3135\\
			&during&$0.1094^{*}$&$0.1690^{*}$&0.0036&0.0000&$0.9882^{*}$&0.5048\\
			\cline{1-8}		
			\hline
		\end{tabular}
		\begin{tablenotes}[para,flushleft]
			\footnotesize
			* Denotes statistical significance at the 5\% level. \\
		\end{tablenotes}
	\end{threeparttable}
\end{table}

\begin{figure}
	\centerline{\includegraphics[width=18cm]{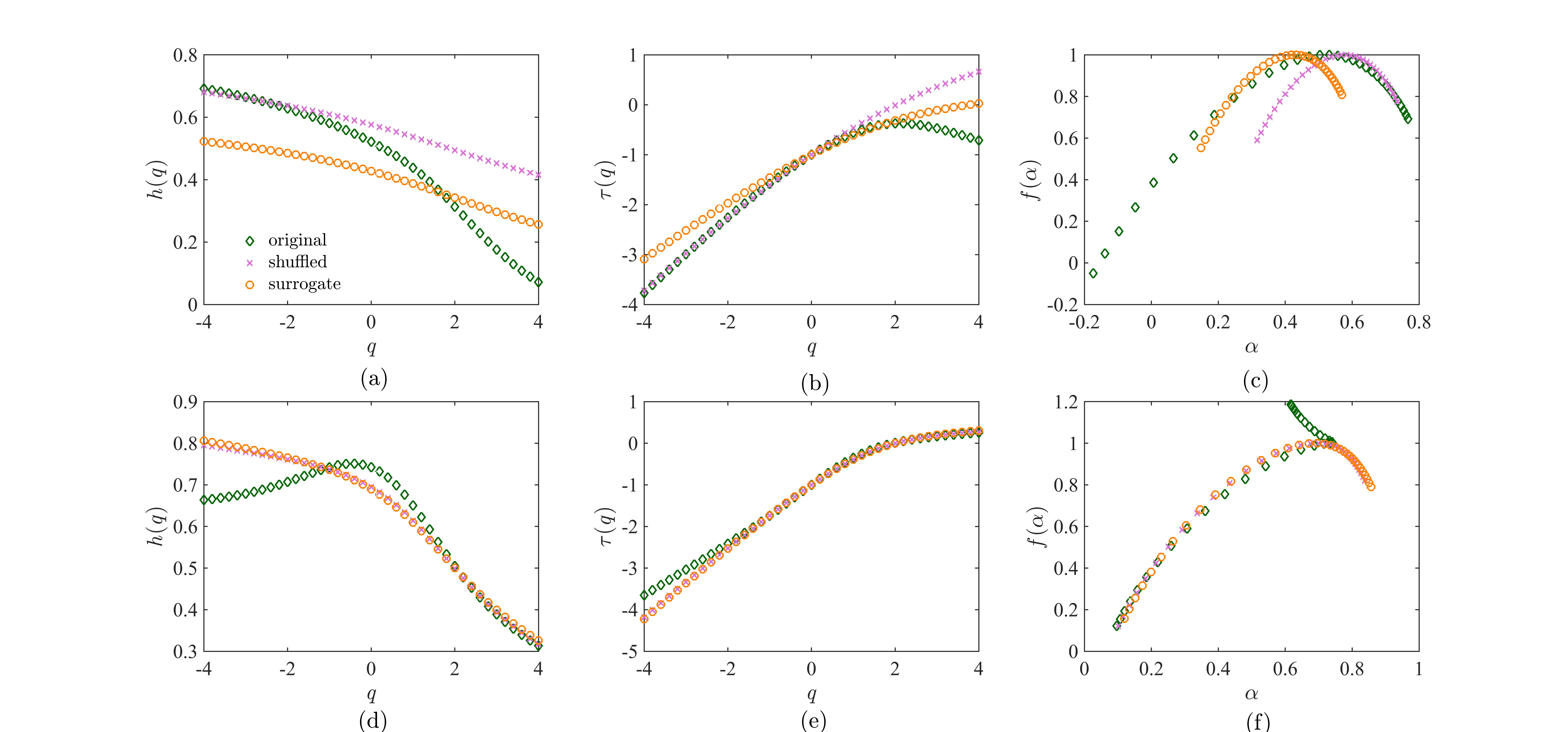}}
	\caption{The multifractal spectrum of SC (a-c) pre-and (d-f) during the COVID-19.}
	\label{Fig:SC}
\end{figure}

To further investigate the impact of COVID-19, we perform MF-DFA to assess degree of SC market efficiency. Specifically, the ranges of the order $q$ and segment size  $s$ depend on the length $N$ of the datasets. we set $q$ from -4 to 4 and $s$ from 30 to one fifth of the subsample length. The ranges of the two parameters are short enough to avoid possible finite size effect and large enough for accurate determination \citep{zhou2012finite}. We illustrate results before and during the pandemic in Fig.~\ref{Fig:SC} and Table~\ref{table:SC}. For a monofractal series, the generalized Hurst exponent $h(q)$ behaves as a constant function of the order $q$ and the mass exponent function $\tau(q)$ depends linearly on $q$. Fig.~\ref{Fig:SC} shows that $h(q)$ of the original SC series during the two periods are decreasing as the order $q$ increases, and there is a nonlinear relationship between $\tau(q)$ and $q$. Consistent with work of \cite{wang2019multifractal}, our results provide evidence of multifractal characteristics in SC. Apart from that, the curves exhibit discrepancy, which indicates different multifractal behavior during the two periods. Moreover,  Table~\ref{table:SC} shows that SC has weaker multifractal feature and smaller $MDM$ during the COVID-19, which means higher market efficiency. 

These results suggest that SC is weak-form efficient during not before the COVID-19 pandemic. Our finding is different from previous study of \cite{yang2020pricing}, which has suggested that SC is close to weak-form efficient. Ref. \cite{yang2020pricing} uses daily data around one year while each of our data sample spans for two months. Thus a possible explanation for the different results might be that a financial market may be efficient in the long run, while sometimes exhibits short-run inefficiencies \citep{yang2019revisiting}. 

\begin{table}[h]
	\renewcommand\arraystretch{1.5}
	\small
	\centering
	\caption{Multifractal strength and market inefficiency of SC.}
	\label{table:SC}
	\begin{tabular}{llllllllll}		\cline{1-10}
		&&\multicolumn{2}{l}{$h(2)$}&\multicolumn{2}{l}{$\Delta h$}& \multicolumn{2}{l}{$\Delta \alpha$}&\multicolumn{2}{l}{$MDM$}\\
		\cline{1-10}
		&&Before&During&Before&During& Before & During& Before & During\\
		\cline{1-10}
		&Original&0.3135&0.5048&0.6199&0.4375&0.9900&0.6448&0.3100&0.1751\\
		&Shuffled&0.4944&0.4993&0.2649&0.4835&0.4186&0.7444&0.1326&0.2411\\
		&Surrogate&0.3422&0.5002&0.2662&0.4807&0.4210&0.7392&0.1405&0.2400\\		\cline{1-10}
	\end{tabular}
\end{table}

\subsection{The sources of multifractality in SC}
\label{s:s:source}
In this section we study sources of multifractality in SC. Generally speaking, multifractal properties in time series mainly come from temporal correlation and non-Gaussian probability distribution \citep{kantelhardt2002multifractal, zhou2009components, oh2012multifractal,zhou2012finite, 2019multifractal}. Firstly, we shuffle the data randomly to remove potential temporal correlation and preserve the original probability distribution. If shuffled data has weaker multifractal strength, both temporal correlation and probability distribution contribute to the multifractality of raw data. If not, other factors are the potential origins of multifractal properties. Secondly, we generate surrogate data via the iterative amplitude adjusted Fourier transform (IAAFT) algorithm \citep{schreiber1996improved} to investigate the impact of nonlinear correlation on multifractality. The surrogate time series has the same linear correlation and probability distribution as the raw data, though the nonlinear correlation is eliminated. If the surrogate data exhibits lower multifractal behavior than the original data, the nonlinear correlation is a source of multifractality in the dataset. Otherwise this factor has little or no effect on the multifractal characteristics of the data.

In Fig.~\ref{Fig:SC} and Table~\ref{table:SC}, the shuffled data before the COVID-19 has narrower multifractal spectrum and smaller $\Delta\alpha$ than the original data. The shuffling procedure doesn't eliminate all the multifractal nature of SC. Accordingly we deduce that non-Gaussian probability distribution contributes to the multifractality in SC during this period. Linear and nonlinear correlations are also possible sources of multifractal nature. Then we generate 50 surrogate data to investigate the respective effects of the two factors. 
Compared with the original time series, the surrogate data also exhibits weaker multifractal strength, which confirms impact of nonlinear correlation on multifractality. Additionally, shuffled data and surrogate time series have similar multifractal properties,  including the generalized Hurst exponent $h(q)$, mass exponent function $\tau(q)$ and multifractal spectrum $f(\alpha)$. Thus we conclude that linear correlation hardly has an impact on multifractal properties in SC. Before the COVID-19, non-Gaussian probability distribution and nonlinear correlations are the two major sources of multifractality in SC.

We find that during the COVID-19 pandemic, shuffled data has a little higher degree of multifractality than the raw time series, which confirms that multifractal character of SC in this period mainly derives from non-Gaussian probability distribution. Temporal correlations have little or no impact on the multifractality of SC. Similarly, compared with the raw data, the  surrogate data has slightly higher multifractal degree. This provides evidence that nonlinear structure has almost no contribution to the multifractal nature of SC. In Fig.~\ref{Fig:SC} and Table~\ref{table:SC}, the shuffled data has almost the same multifractality as the surrogate data, which supports no effect of linear structure on multifractal nature in SC. During the COVID-19 pandemic, the non-Gaussian probability distribution is the major source of multifractality of SC. The present results differ from study  of \cite{wang2019multifractal}. A possible explanation might be that we use different sample periods and time frequency from \cite{wang2019multifractal}. 

We conclude that the components of multifractality on SC vary before and during the pandemic. Nonlinear correlation and non Gaussian probability distribution are the main sources of multifractality in SC before the COVID-19, while during the pandemic non-Gaussian probability distribution contributes to the multifractal character. This change indicates the impact of  COVID-19 on SC.

\subsection{DCCA coefficient}
\begin{figure}[h]
	\centering
	\centerline{\includegraphics[width=15cm]{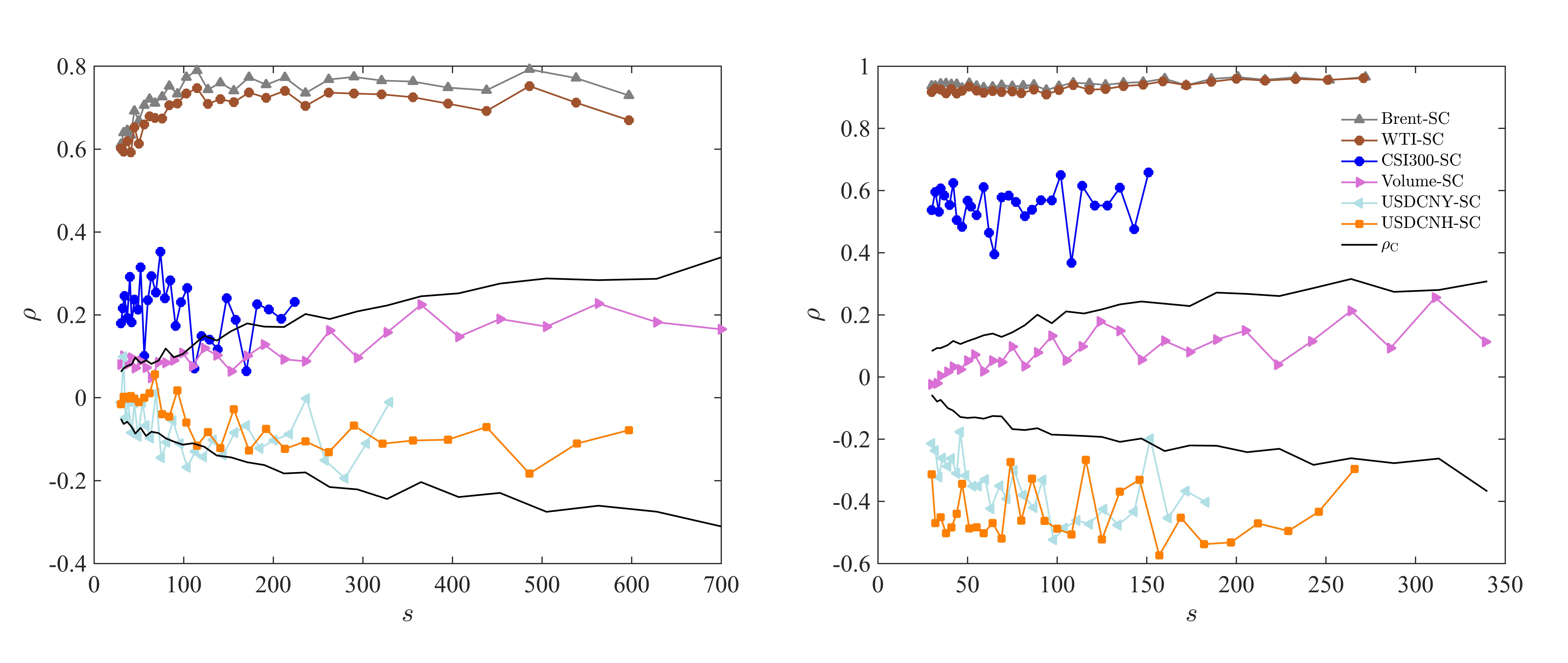}}	
	\caption{DCCA coefficients $\rho_{DCCA}$ between SC and other time series (a) pre-and (b) during the COVID-19. The black lines represent the critical values $\rho_{c}$ for the 95\% confidence level under assumption of no cross-correlations.}
	\label{Fig:rou}
\end{figure}

We illustrate $\rho_{DCCA}$ between SC and other financial assets returns as well as SC trading volume changes in Fig.~\ref{Fig:rou}. There are significant positive cross-correlations between SC and CSI300 along with the two global crude oil benchmarks for almost all time scales before the pandemic. The correlation between SC and Chinese stock market is also observed in previous study \citep{zhu2021multidimensional}. Similarly, our results provide evidence of linkages between SC and the two international major crude oil markets, which is in accordance with work of \cite{yang2020return} and \cite{yang2020pricing}. SC has no significant cross-correlations with onshore and offshore RMB exchange rates, which is consistent with study of \cite{lin2021china}. Besides that, our finding differs from previous study \citep{zhang2021cross} which suggests the correlation between SC price and volume. A possible explanation for this might be the different sample periods and time frequencies. 

Fig.~\ref{Fig:rou}b shows that during the COVID-19, SC has significant cross-correlations with all the other series except for the trading volume. What can be clearly seen in Fig.~\ref{Fig:rou}b  is the increase of $\rho_{DCCA}$ between SC returns and other financial assets returns. The values of $\rho_{DCCA}$ between SC and the two international crude oil benchmarks are extremely close to 1 for most time scales, which indicates perfect positive cross-correlations. And interestingly, SC is negatively cross-correlated with the onshore and offshore exchange rates during but not before the COVID-19, which also implies that SC is more connected with other financial markets after the outbreak of the COVID-19 pandemic. As a result, these stronger co-movements increased symmetric risks to a great extent.

\subsection{Multifractal cross-correlation between SC and other financial markets}
\label{s:s:MFDCCA}
\begin{figure}
	\centerline{\includegraphics[width=17cm]{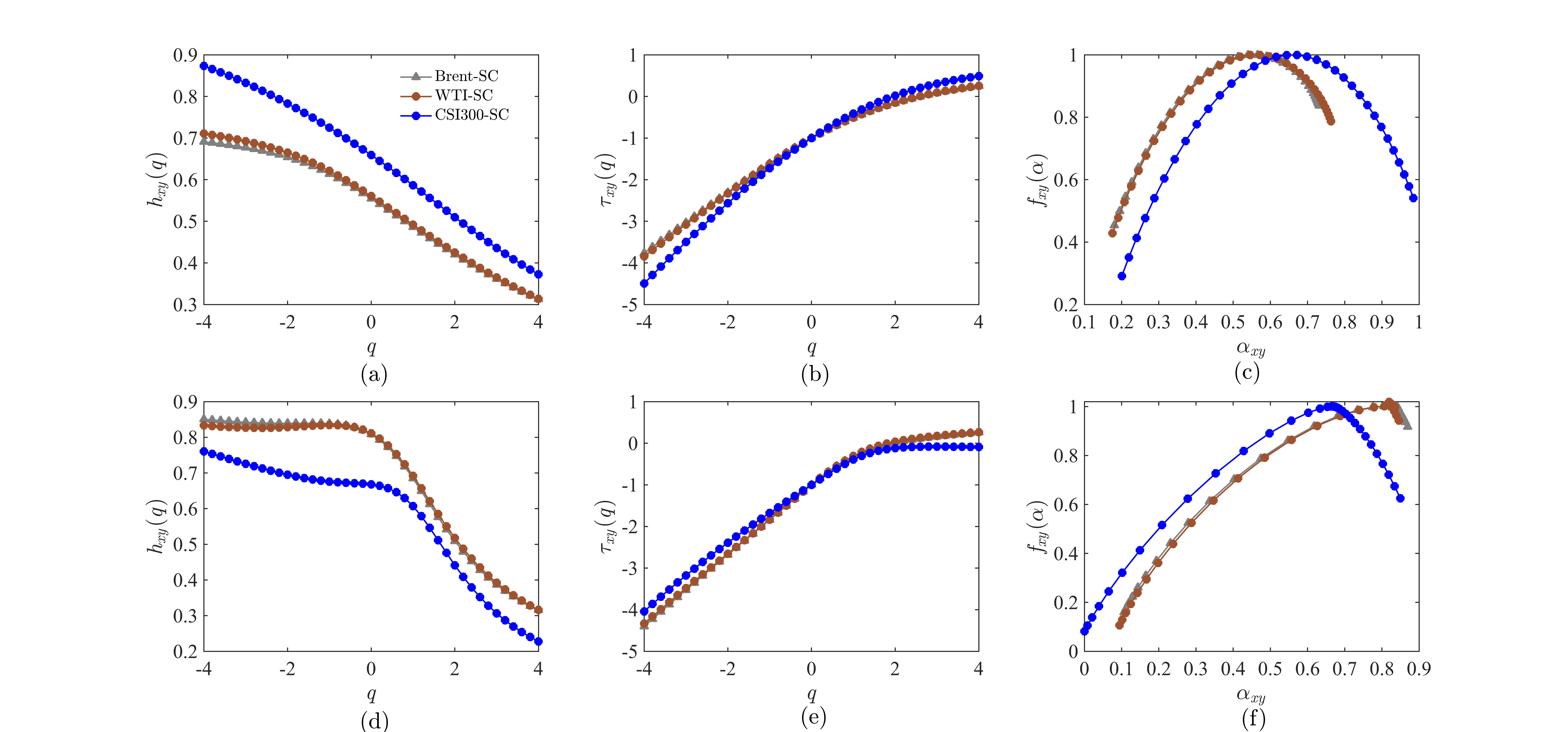}}
	\caption{Multifractal cross-correlation analysis of SC and other assets returns (a-c) before and (d-f) during the COVID-19 pandemic. From left to right, each column corresponds to $h_{xy}(q)$, $\tau_{xy}(q)$ and $f_{xy}(\alpha)$.}
	\label{Fig:X}
\end{figure}

We apply the MF-DCCA to measure the multifractal cross-correlations between SC and the Chinese stock market as well as other crude oil futures markets. Fig.~\ref{Fig:X} and Table~\ref {table:X} show the multifractal detrended cross-correlation analysis of SC and other financial instruments before and during the pandemic. Fig.~\ref{Fig:X}a, d illustrate the evolution of the generalized bivariate Hurst exponents $h_{xy}(q)$ and the order $q$. During the two periods, all the scaling exponents $h_{xy}(q)$ vary with the order $q$. Fig.~\ref{Fig:X}b and e show that all the joint mass exponents $\tau_{xy}(q)$ have a nonlinear relationships with the order $q$, which also indicates the multifractal cross-correlation between SC and other financial markets. Fig.~\ref{Fig:X}c and f depict the correlation curves between singularity strength $\alpha_{xy}$ and multifractal spectrum $f_{xy}(\alpha)$. We observe that the multifractal cross-correlations between SC and WTI are extremely similar to that between SC and Brent. This result may be explained by the fact that WTI has an extremely close connection with Brent \citep{yang2020time}. 

\begin{table}[h]
	\renewcommand\arraystretch{1.5}
	\small
	\centering
	\caption{Multifractal cross-correlation between SC and other markets.}
	\label{table:X}
	\begin{tabular}{llllllll}		\cline{1-8}
		&&\multicolumn{2}{l}{$h_{xy}(2)$}&\multicolumn{2}{l}{$\Delta h_{xy}$}& \multicolumn{2}{l}{$\Delta \alpha_{xy}$}\\
		\cline{1-8}
		&&Before&During&Before&During& Before & During\\
		\cline{1-8}
		&Brent-SC&0.4201&0.5091&0.3782&0.5347&0.5508&0.7627\\
		&WTI-SC&0.4249&0.5178&0.3971&0.5170&0.5876&0.7510\\
		&CSI300-SC&0.5098&0.4413&0.5009&0.5329&0.7842&0.8547\\
		\cline{1-8}
	\end{tabular}
\end{table}

Table~\ref {table:X} reveals that before the COVID-19 pandemic, $h_{xy}(2)$ of Brent-SC and WTI-SC is 0.4201 and 0.4249 respectively, which suggests that SC has similarly long-range cross-correlations with the two international crude oil benchmarks. SC is short-range cross-correlated with CSI300 ($h_{xy}(2)$=0.5098). SC and the Chinese stock market has larger $\Delta h_{xy}$ and $\Delta\alpha_{xy}$ than those of other pairs, which indicates that SC shows strongest multifractal cross-correlation with the Chinese stock market. 

During the COVID-19 pandemic, SC is long-range cross-correlated with CSI300 ($h_{xy}(2)$=0.4413) but short-range cross-correlated with the two international crude oil benchmarks ($h_{xy}(2)$ are 0.5091 and 0.5178 respectively). And SC presents similar multifractal cross-correlations with the two international crude oil benchmarks. The multifractal cross-correlation between SC and CSI300 is stronger than that between SC and the two international crude oil benchmarks. And SC has stronger multifractal cross-correlations with other three financial instruments in this periods, which also indicates the influence of COVID-19 on the Chinese crude oil futures market.

\section{Conclusion}
\label{s:conclusion}
In this paper, we explore the short-term impact of COVID-19 on SC. We compare the market efficiency and multifractal nature of SC before and during this pandemic, as well as the multifractal cross-correlations with other financial assets returns and SC trading volume changes.

First, we examine market efficiency of SC via the MF-DFA and other commonly used tests to obtain reliable results. We find that the Chinese crude oil futures market is close to weak-form efficient during but not before the pandemic. The improvement of SC market efficiency may partly be explained by the sample periods. To explore the short-term effects of the pandemic on SC, we select the sub-sample periods of two months. If the sample period is too long, it will tend to be too noisy for reliable analysis. And previous studies demonstrate that crude oil market is weak-form efficient in the long term \citep{shao2020does,yang2020pricing,kristoufek2014commodity,ghazani2019testing}, though it may show efficiency and inefficiency alternatively at short time scales due to various factors such as policy shift and geopolitical events \citep{Shao2019,yang2020time}. Moreover, during the COVID-19 pandemic period, INE launched plentiful restrictions such as trading hours adjustment to calm investors and support markets. These factors might improve market efficiency temporarily.    

Then we apply the MF-DFA to investigate the sources of multifractal characteristics in SC. Results of numerical simulation show that the nonlinear correlation and non-Gaussian probability distribution contribute to the multifractal characteristics of SC before the COVID-19, while the non-Gaussian probability distribution is the major source during this pandemic. These changes of multifractal sources also provide evidence for the impact of the new coronavirus on SC.

Third, we use the DCCA coefficients to determine whether the cross-correlations between SC returns and other financial assets returns as well as trading volume changes exist. Then we utilize MF-DCCA to quantify the multifractal cross-correlations. Before the COVID-19, there are significant cross-correlations between SC and the two international benchmarks as well as the Chinese stock market. During the pandemic, SC has significant cross-correlations with all the returns even including the onshore and offshore exchange rates. Throughout the sampling periods, there is no significant cross-correlation between SC and its trading volume. The strong correlation between SC and CSI300 may be explained by the fact that the social and economic environments are relatively homogenous within the same country. Besides that, SC shows higher cross-correlations with all the other financial returns in this period than before. The stronger links between SC and other financial markets are likely to be related to the COVID-19 pandemic. It has been suggested that co-movements of financial markets increased significantly during extremely periods like the 2008 financial crisis \citep{yang2020time} and COVID-19 pandemic \citep{li2021return}. The rapidly expanding coronavirus crisis triggered turmoil among global financial markets and investors, which brought an unprecedented level of risk and volatility to assets. Again, sub-sample periods of two-months are too short to find effective coronavirus treatments and vaccines. Various interventions such as travel restrictions launched by many governments to prevent the transmission of COVID-19 were not yet fully effective during this sample period. Consequently, the pandemic increased connectedness across global financial markets and lowered likelihood of diversification benefits greatly. 

Our findings confirm that the COVID-19 has a significant short-term impact on the market efficiency and multifractality of the Chinese crude oil futures. The COVID-19 pandemic increased correlation risk and reduced arbitrage opportunities between SC and other financial markets remarkably. Additionally, given the continuous spread of COVID-19 globally, market participants should improve their risk assessment to minimize possible losses triggered by the COVID-19. And the SC investors should also implement adequate strategies to manage the accompanying uncertainty. Policymakers should focus on alleviating the panic and increasing traders' confidence in market recovery. In addition, several questions still remain to be answered. In this study, we mainly focus on the short-term effects of COVID-19 on SC. A further study could assess the long-lasting influence of this pandemic on the major crude oil futures markets including SC. 

\section*{Acknowledgements}
This work was supported by the National Natural Science Foundation of China (12005064, 11805119).


\section*{References}
\begin{thebibliography}{10}
	\expandafter\ifx\csname url\endcsname\relax
	\def\url#1{\texttt{#1}}\fi
	\expandafter\ifx\csname urlprefix\endcsname\relax\def\urlprefix{URL }\fi
	\expandafter\ifx\csname href\endcsname\relax
	\def\href#1#2{#2} \def\path#1{#1}\fi
	
	\bibitem{yang2020pricing}
	C.~Yang, F.~Lv, L.-B. Fang, X.-X. Shang, {The pricing efficiency of crude oil
		futures in the Shanghai International Exchange}, Financ. Res. Lett. 36 (2020)
	101329.
	\newblock \href {http://dx.doi.org/10.1016/j.frl.2019.101329}
	{\path{doi:10.1016/j.frl.2019.101329}}.
	
	\bibitem{lu2020examining}
	X.-J. Lu, F.~Ma, J.-Q. Wang, J.-Q. Wang, {Examining the predictive information
		of CBOE OVX on China's oil futures volatility: Evidence from MS-MIDAS
		models}, Energy 212 (2020) 118743.
	\newblock \href {http://dx.doi.org/10.1016/j.energy.2020.118743}
	{\path{doi:10.1016/j.energy.2020.118743}}.
	
	\bibitem{zhang2021exploring}
	Y.-J. Zhang, S.-J. Ma, {Exploring the dynamic price discovery, risk transfer
		and spillover among INE, WTI and Brent crude oil futures markets: Evidence
		from the high-frequency data}, Int. J. Financ. \& Econ. 26~(2) (2021)
	2414--2435.
	\newblock \href {http://dx.doi.org/10.1002/ijfe.1914}
	{\path{doi:10.1002/ijfe.1914}}.
	
	\bibitem{zhang2021cross}
	S.-C. Zhang, Y.-Q. Guo, H.~Cheng, H.-W. Zhang, {Cross-correlations between
		price and volume in China's crude oil futures market: A study based on
		multifractal approaches}, Chaos, Solitons \& Fractals 144 (2021) 110642.
	\newblock \href {http://dx.doi.org/10.1016/j.chaos.2020.110642}
	{\path{doi:10.1016/j.chaos.2020.110642}}.
	
	\bibitem{yang2020time}
	Y.-H. Yang, Y.-H. Shao, H.-L. Shao, X.~Song, {The time-dependent lead-lag
		relationship between WTI and Brent crude oil spot markets}, Front. Phys. 8
	(2020) 132.
	\newblock \href {http://dx.doi.org/10.3389/fphy.2020.00132}
	{\path{doi:10.3389/fphy.2020.00132}}.
	
	\bibitem{shao2020does}
	Y.-H. Shao, {Does crude oil market efficiency improve after the lift of the US
		export ban? Evidence from time-varying Hurst exponent}, Front. Phys. 8 (2020)
	405.
	\newblock \href {http://dx.doi.org/10.3389/fphy.2020.551501}
	{\path{doi:10.3389/fphy.2020.551501}}.
	
	\bibitem{kantelhardt2002multifractal}
	J.~W. Kantelhardt, S.~A. Zschiegner, E.~Koscielny-Bunde, S.~Havlin, A.~Bunde,
	H.~E. Stanley, {Multifractal detrended fluctuation analysis of nonstationary
		time series}, Phys. A Stat. Mech. Appl. 316~(1-4) (2002) 87--114.
	\newblock \href {http://dx.doi.org/10.1016/S0378-4371(02)01383-3}
	{\path{doi:10.1016/S0378-4371(02)01383-3}}.
	
	\bibitem{zhou2008multifractal}
	W.-X. Zhou, {Multifractal detrended cross-correlation analysis for two
		nonstationary signals}, Phys. Rev. E 77~(6) (2008) 66211.
	\newblock \href {http://dx.doi.org/10.1103/PhysRevE.77.066211}
	{\path{doi:10.1103/PhysRevE.77.066211}}.
	
	\bibitem{2019multifractal}
	Z.-Q. Jiang, W.-J. Xie, W.-X. Zhou, D.~Sornette, {Multifractal analysis of
		financial markets: a review}, Reports Prog. Phys. 82~(12) (2019) 125901.
	\newblock \href {http://dx.doi.org/10.1088/1361-6633/ab42fb}
	{\path{doi:10.1088/1361-6633/ab42fb}}.
	
	\bibitem{kristoufek2014commodity}
	L.~Kristoufek, M.~Vosvrda, {Commodity futures and market efficiency}, Energy
	Econ. 42 (2014) 50--57.
	\newblock \href {http://dx.doi.org/10.1016/j.eneco.2013.12.001}
	{\path{doi:10.1016/j.eneco.2013.12.001}}.
	
	\bibitem{ghazani2019testing}
	M.~M. Ghazani, S.~B. Ebrahimi, {Testing the adaptive market hypothesis as an
		evolutionary perspective on market efficiency: Evidence from the crude oil
		prices}, Financ. Res. Lett. 30 (2019) 60--68.
	\newblock \href {http://dx.doi.org/10.1016/j.frl.2019.03.032}
	{\path{doi:10.1016/j.frl.2019.03.032}}.
	
	\bibitem{yang2020return}
	J.~Yang, Y.-G. Zhou, {Return and volatility transmission between China's and
		international crude oil futures markets: A first look}, J. Futur. Mark.
	40~(6) (2020) 860--884.
	\newblock \href {http://dx.doi.org/10.1002/fut.22103}
	{\path{doi:10.1002/fut.22103}}.
	
	\bibitem{wang2019multifractal}
	F.~Wang, X.~Ye, C.-X. Wu, {Multifractal characteristics analysis of crude oil
		futures prices fluctuation in China}, Phys. A Stat. Mech. Appl. 533 (2019)
	122021.
	\newblock \href {http://dx.doi.org/10.1016/j.physa.2019.122021}
	{\path{doi:10.1016/j.physa.2019.122021}}.
	
	\bibitem{huang2020identifying}
	X.-H. Huang, S.-P. Huang, {Identifying the comovement of price between China's
		and international crude oil futures: A time-frequency perspective}, Int. Rev.
	Financ. Anal. 72 (2020) 101562.
	\newblock \href {http://dx.doi.org/10.1016/j.irfa.2020.101562}
	{\path{doi:10.1016/j.irfa.2020.101562}}.
	
	\bibitem{ji2021intra}
	Q.~Ji, D.-Y. Zhang, Y.-Q. Zhao, {Intra-day co-movements of crude oil futures:
		China and the international benchmarks}, Ann. Oper. Res.\href
	{http://dx.doi.org/10.1007/s10479-021-04097-x}
	{\path{doi:10.1007/s10479-021-04097-x}}.
	
	\bibitem{zhu2021relationships}
	P.-F. Zhu, Y.~Tang, Y.~Wei, Y.-M. Dai, T.-T. Lu, {Relationships and portfolios
		between oil and Chinese stock sectors: A study based on wavelet
		denoising-higher moments perspective}, Energy 217 (2021) 119416.
	\newblock \href {http://dx.doi.org/10.1016/j.energy.2020.119416}
	{\path{doi:10.1016/j.energy.2020.119416}}.
	
	\bibitem{lv2020crude}
	F.~Lv, C.~Yang, L.~Fang, {Do the crude oil futures of the Shanghai
		International Energy Exchange improve asset allocation of Chinese
		petrochemical-related stocks?}, Int. Rev. Financ. Anal. 71 (2020) 101537.
	\newblock \href {http://dx.doi.org/10.1016/j.irfa.2020.101537}
	{\path{doi:10.1016/j.irfa.2020.101537}}.
	
	\bibitem{mensi2017oil}
	W.~Mensi, S.~Hammoudeh, S.~J.~H. Shahzad, K.~H. Al-Yahyaee, M.~Shahbaz, {Oil
		and foreign exchange market tail dependence and risk spillovers for MENA,
		emerging and developed countries: VMD decomposition based copulas}, Energy
	Econ. 67 (2017) 476--495.
	\newblock \href {http://dx.doi.org/10.1016/j.eneco.2017.08.036}
	{\path{doi:10.1016/j.eneco.2017.08.036}}.
	
	\bibitem{ahmad2020intraday}
	W.~Ahmad, R.~Prakash, G.~S. Uddin, R.~J.~K. Chahal, M.~L. Rahman, A.~Dutta, {On
		the intraday dynamics of oil price and exchange rate: What can we learn from
		China and India?}, Energy Econ. 91 (2020) 104871.
	\newblock \href {http://dx.doi.org/10.1016/j.eneco.2020.104871}
	{\path{doi:10.1016/j.eneco.2020.104871}}.
	
	\bibitem{lin2021china}
	B.-Q. Lin, T.~Su, {Do China's macro-financial factors determine the Shanghai
		crude oil futures market?}, Int. Rev. Financ. Anal. 78 (2021) 101953.
	\newblock \href {http://dx.doi.org/10.1016/j.irfa.2021.101953}
	{\path{doi:10.1016/j.irfa.2021.101953}}.
	
	\bibitem{zhu2021multidimensional}
	P.-F. Zhu, Y.~Tang, Y.~Wei, T.-T. Lu, {Multidimensional risk spillovers among
		crude oil, the US and Chinese stock markets: Evidence during the COVID-19
		epidemic}, Energy 231 (2021) 120949.
	\newblock \href {http://dx.doi.org/10.1016/j.energy.2021.120949}
	{\path{doi:10.1016/j.energy.2021.120949}}.
	
	\bibitem{niu2021role}
	Z.-B. Niu, Y.-Y. Liu, W.~Gao, H.-W. Zhang, {The role of coronavirus news in the
		volatility forecasting of crude oil futures markets: Evidence from China},
	Resour. Policy 73 (2021) 102173.
	\newblock \href {http://dx.doi.org/10.1016/j.resourpol.2021.102173}
	{\path{doi:10.1016/j.resourpol.2021.102173}}.
	
	\bibitem{li2021return}
	X.-F. Li, B.~Li, G.-W. Wei, L.~Bai, Y.~Wei, C.~Liang, {Return connectedness
		among commodity and financial assets during the COVID-19 pandemic: Evidence
		from China and the US}, Resour. Policy 73 (2021) 102166.
	\newblock \href {http://dx.doi.org/10.1016/j.resourpol.2021.102166}
	{\path{doi:10.1016/j.resourpol.2021.102166}}.
	
	\bibitem{wald1940test}
	A.~Wald, J.~Wolfowitz, {On a test whether two samples are from the same
		population}, Ann. Math. Stat. 11~(2) (1940) 147--162.
	\newblock \href {http://dx.doi.org/10.1214/aoms/1177731909}
	{\path{doi:10.1214/aoms/1177731909}}.
	
	\bibitem{ljung1978measure}
	G.~M. Ljung, G.~E.~P. Box, {On a measure of lack of fit in time series models},
	Biometrika 65~(2) (1978) 297--303.
	\newblock \href {http://dx.doi.org/10.1093/biomet/65.2.297}
	{\path{doi:10.1093/biomet/65.2.297}}.
	
	\bibitem{lo1988stock}
	A.~W. Lo, A.~C. MacKinlay, {Stock Market Prices Do Not Follow Random Walks:
		Evidence from a Simple Specification Test}, Rev. Financ. Stud. 1~(1) (1988)
	41--66.
	\newblock \href {http://dx.doi.org/10.1093/rfs/1.1.41}
	{\path{doi:10.1093/rfs/1.1.41}}.
	
	\bibitem{broock1996test}
	W.~A. Broock, J.~A. Scheinkman, W.~D. Dechert, B.~LeBaron, {A test for
		independence based on the correlation dimension}, Econom. Rev. 15~(3) (1996)
	197--235.
	\newblock \href {http://dx.doi.org/10.1080/07474939608800353}
	{\path{doi:10.1080/07474939608800353}}.
	
	\bibitem{mann1945nonparametric}
	H.~B. Mann, {Nonparametric tests against trend}, Econom. J. Econom. Soc. 13~(3)
	(1945) 245--259.
	\newblock \href {http://dx.doi.org/10.2307/1907187}
	{\path{doi:10.2307/1907187}}.
	
	\bibitem{kendall1975}
	M.~G. Kendall, {Rank Correlation Methods}, 4th Edition, Griffin, London, 1975.
	
	\bibitem{peng1994mosaic}
	C.-K. Peng, S.~V. Buldyrev, S.~Havlin, M.~Simons, H.~E. Stanley, A.~L.
	Goldberger, {Mosaic organization of DNA nucleotides}, Phys. Rev. E 49~(2)
	(1994) 1685.
	\newblock \href {http://dx.doi.org/10.1103/PhysRevE.49.1685}
	{\path{doi:10.1103/PhysRevE.49.1685}}.
	
	\bibitem{podobnik2011statistical}
	B.~Podobnik, Z.-Q. Jiang, W.-X. Zhou, H.~E. Stanley, {Statistical tests for
		power-law cross-correlated processes}, Phys. Rev. E 84~(6) (2011) 66118.
	\newblock \href {http://dx.doi.org/10.1103/PhysRevE.84.066118}
	{\path{doi:10.1103/PhysRevE.84.066118}}.
	
	\bibitem{halsey1986fractal}
	T.~C. Halsey, M.~H. Jensen, L.~P. Kadanoff, I.~Procaccia, B.~I. Shraiman,
	{Fractal measures and their singularities: The characterization of strange
		sets}, Phys. Rev. A 33~(2) (1986) 1141.
	\newblock \href {http://dx.doi.org/10.1103/PhysRevA.33.1141}
	{\path{doi:10.1103/PhysRevA.33.1141}}.
	
	\bibitem{shao2021Fractals}
	Y.-H. Shao, H.~Xu, Y.-L. Liu, H.-C. Xu, {Multifractal behavior of
		cryptocurrencies before and during COVID-19}, Fractals 29~(06) (2021)
	2150132.
	\newblock \href {http://dx.doi.org/10.1142/S0218348X21501322}
	{\path{doi:10.1142/S0218348X21501322}}.
	
	\bibitem{Shao-Gu-Jiang-Zhou-Sornette-2012-SR}
	Y.-H. Shao, G.-F. Gu, Z.-Q. Jiang, W.-X. Zhou, D.~Sornette, {Comparing the
		performance of FA, DFA and DMA using different synthetic long-range
		correlated time series}, Sci. Rep. 2 (2012) 835.
	\newblock \href {http://dx.doi.org/10.1038/srep00835}
	{\path{doi:10.1038/srep00835}}.
	
	\bibitem{shao2015effects}
	Y.-H. Shao, G.-F. Gu, Z.-Q. Jiang, W.-X. Zhou, {Effects of polynomial trends on
		detrending moving average analysis}, Fractals 23~(03) (2015) 1550034.
	\newblock \href {http://dx.doi.org/10.1142/S0218348X15500346}
	{\path{doi:10.1142/S0218348X15500346}}.
	
	\bibitem{lim2011evolution}
	K.-P. Lim, R.~Brooks, {The evolution of stock market efficiency over time: A
		survey of the empirical literature}, J. Econ. Surv. 25~(1) (2011) 69--108.
	\newblock \href {http://dx.doi.org/10.1111/j.1467-6419.2009.00611.x}
	{\path{doi:10.1111/j.1467-6419.2009.00611.x}}.
	
	\bibitem{yang2019revisiting}
	Y.-H. Yang, Y.-H. Shao, H.-L. Shao, H.~E. Stanley, {Revisiting the weak-form
		efficiency of the EUR/CHF exchange rate market: Evidence from episodes of
		different Swiss franc regimes}, Phys. A Stat. Mech. Appl. 523 (2019)
	734--746.
	\newblock \href {http://dx.doi.org/10.1016/j.physa.2019.02.056}
	{\path{doi:10.1016/j.physa.2019.02.056}}.
	
	\bibitem{zunino2008multifractal}
	L.~Zunino, B.~M. Tabak, A.~Figliola, D.~G. P{\'{e}}rez, M.~Garavaglia, O.~A.
	Rosso, {A multifractal approach for stock market inefficiency}, Phys. A Stat.
	Mech. Appl. 387~(26) (2008) 6558--6566.
	\newblock \href {http://dx.doi.org/10.1016/j.physa.2008.08.028}
	{\path{doi:10.1016/j.physa.2008.08.028}}.
	
	\bibitem{wang2009analysis}
	Y.-D. Wang, L.~Liu, R.-B. Gu, {Analysis of efficiency for Shenzhen stock market
		based on multifractal detrended fluctuation analysis}, Int. Rev. Financ.
	Anal. 18~(5) (2009) 271--276.
	\newblock \href {http://dx.doi.org/10.1016/j.irfa.2009.09.005}
	{\path{doi:10.1016/j.irfa.2009.09.005}}.
	
	\bibitem{podobnik2008detrended}
	B.~Podobnik, H.~E. Stanley, Detrended cross-correlation analysis: a new method
	for analyzing two nonstationary time series, Phys. Rev. Lett. 100~(8) (2008)
	084102.
	\newblock \href {http://dx.doi.org/10.1103/PhysRevLett.100.084102}
	{\path{doi:10.1103/PhysRevLett.100.084102}}.
	
	\bibitem{zebende2011dcca}
	G.~F. Zebende, {DCCA cross-correlation coefficient: Quantifying level of
		cross-correlation}, Phys. A Stat. Mech. Appl. 390~(4) (2011) 614--618.
	\newblock \href {http://dx.doi.org/10.1016/j.physa.2010.10.022}
	{\path{doi:10.1016/j.physa.2010.10.022}}.
	
	\bibitem{wang2013random}
	G.-J. Wang, C.~Xie, S.~Chen, J.-J. Yang, M.-Y. Yang, {Random matrix theory
		analysis of cross-correlations in the US stock market: Evidence from
		Pearson’s correlation coefficient and detrended cross-correlation
		coefficient}, Phys. A Stat. Mech. Appl. 392~(17) (2013) 3715--3730.
	\newblock \href {http://dx.doi.org/10.1016/j.physa.2013.04.027}
	{\path{doi:10.1016/j.physa.2013.04.027}}.
	
	\bibitem{zhou2012finite}
	W.-X. Zhou, {Finite-size effect and the components of multifractality in
		financial volatility}, Chaos, Solitons \& Fractals 45~(2) (2012) 147--155.
	\newblock \href {http://dx.doi.org/10.1016/j.chaos.2011.11.004}
	{\path{doi:10.1016/j.chaos.2011.11.004}}.
	
	\bibitem{zhou2009components}
	W.-X. Zhou, {The components of empirical multifractality in financial returns},
	Europhys. Lett. 88~(2) (2009) 28004.
	\newblock \href {http://dx.doi.org/10.1209/0295-5075/88/28004}
	{\path{doi:10.1209/0295-5075/88/28004}}.
	
	\bibitem{oh2012multifractal}
	G.~Oh, C.~Eom, S.~Havlin, W.-S. Jung, F.~Wang, H.~E. Stanley, S.~Kim, {A
		multifractal analysis of Asian foreign exchange markets}, Eur. Phys. J. B
	85~(6) (2012) 214.
	\newblock \href {http://dx.doi.org/10.1140/epjb/e2012-20570-0}
	{\path{doi:10.1140/epjb/e2012-20570-0}}.
	
	\bibitem{schreiber1996improved}
	T.~Schreiber, A.~Schmitz, {Improved surrogate data for nonlinearity tests},
	Phys. Rev. Lett. 77~(4) (1996) 635.
	\newblock \href {http://dx.doi.org/10.1103/PhysRevLett.77.635}
	{\path{doi:10.1103/PhysRevLett.77.635}}.
	
	\bibitem{Shao2019}
	Y.-H. Shao, Y.-H. Yang, H.-L. Shao, H.~E. Stanley, {Time-varying lead – lag
		structure between the crude oil spot and futures markets}, Phys. A Stat.
	Mech. Appl. 523 (2019) 723--733.
	\newblock \href {http://dx.doi.org/10.1016/j.physa.2019.03.002}
	{\path{doi:10.1016/j.physa.2019.03.002}}.
	
\end{thebibliography}
\end{document}